\documentclass[aps,prl,amssymb,showpacs,superscriptaddress,twocolumn]{revtex4}
\usepackage{graphicx}

\newcommand{\Case}[2]{{\textstyle \frac{#1}{#2}}}
\newcommand{\lP}{\ell_{\mathrm P}}

\newcommand{\sgn}{\mathop{\mathrm{sgn}}}

\begin{document}
%

%\today

\preprint{IMSc/2004/07/28}

%\title{Absence of particle horizon in semi-classical Loop Quantum Cosmology}
\title{Genericness of inflation in isotropic loop quantum cosmology}

\author{Ghanashyam Date}
\email{shyam@imsc.res.in}
\affiliation{The Institute of Mathematical Sciences\\
CIT Campus, Chennai-600 113, INDIA.}

\author{Golam Mortuza Hossain}
\email{golam@imsc.res.in}
\affiliation{The Institute of Mathematical Sciences\\
CIT Campus, Chennai-600 113, INDIA.}

\begin{abstract}
Non-perturbative corrections from loop quantum cosmology (LQC) to the 
scalar matter sector is already known to imply inflation. We prove that
the LQC modified scalar field generates exponential inflation in the
small scale factor regime, for all positive definite potentials, independent
of initial conditions and independent of ambiguity parameters. For positive
semi-definite potentials it is always possible to choose, without fine
tuning, a value of one of the ambiguity parameters such that exponential
inflation results, provided zeros of the potential are approached at most
as a power law in the scale factor. In conjunction with generic occurrence 
of bounce at small volumes, particle horizon is absent thus eliminating
the horizon problem of the standard Big Bang model.

\end{abstract}

\pacs{04.60.Pp, 04.60.Kz, 98.80.Jk}

\maketitle

%%%%%%%%%%%%%%%%%%%%%%%%%%%%%%%%%%%%%%%%%%%%%%%%%%%%%%%%%%%%%%%%%%%%%%%

The Standard Big Bang model so far is the most successful large scale 
description of our universe. In this description, the evolution of our 
universe begins from a singularity. Within the context of homogeneous 
and isotropic expanding space-times, the singularity is unavoidable as 
long as the matter satisfies the so called strong energy condition. The
singularity in this context means that the scale factor (or size of the
universe) vanishes a finite time ago. This vanishing size also implies
that the energy density diverges at this time. Furthermore, the scale
factor vanishes slower than linearly with the synchronous time making
the conformal time integral finite thus implying the existence of particle
horizon.

The particle horizon with respect to a space-time point is defined by the
maximum proper distance a particle could have traveled since the 
beginning of the universe. Due to the behavior of the scale factor, this
is a finite distance. It also means that any space-time point could have 
causal contact only with a finite patch of the space-time around it. By
itself, existence of particle horizon need not be a problem. However, in
conjunction with the thermal history of the universe, the finite horizon
size implies that the surface of last scatter of the cosmic microwave
background photons has regions which could not have been in causal
contact. Yet, there is remarkable isotropy (to within few parts of
hundred thousand) in their angular distribution. This is the {\em horizon
problem} of the Big Bang model.

The most popular approach to solve this puzzle (along with few other 
puzzles) is to introduce a phase of {\em inflation} \cite{GuthInflation}.
Phase of inflation generally refers to a period during which the universe 
goes through a rapid (generally exponential) expansion. Clearly there must 
be a violation of the strong energy condition during inflation. This is 
generally achieved by introducing a scalar field (an inflaton) with a self
interaction potential. By now there are several versions of inflationary 
models \cite{InflationRev}. Generically these solve the horizon problem 
(and other traditional problems such as the flatness problem) and in addition 
make specific predictions about the power spectra of inhomogeneous 
perturbations. While these are attractive features of inflationary models,
generally they need fine tuning the potential and initial conditions for the
inflaton to ensure a sufficient amount of inflation with graceful exit. In 
a sense, the isotropic singularity in Einsteinian gravity implies existence 
of particle horizon which leads to the horizon problem which needs an 
inflationary scenario to be implanted. 

The space-time singularity, however, signals breakdown of the theoretical
framework of classical general relativity. It is widely expected that a
quantum theory of gravity will provide a more accurate description which
will hopefully be free of such breakdowns. A fully satisfactory quantum 
theory of gravity is not yet available. Over the past couple of decades, 
two promising approaches have emerged, String Theory \cite{StringRef} and 
Loop Quantum Gravity (LQG) \cite{LQGRev}. In the last few years, a detailed
adaptation of LQG methods to cosmological context has been developed and has
come to be known as Loop Quantum Cosmology (LQC) \cite{LQCRev}. In this 
letter we work within the LQC framework \cite{Bohr} and more specifically 
within the context of spatially flat or close isotropic models.

It has already been shown that the LQC framework is free of singularity,
both in the isotropic context \cite{Sing} as well as more generally for 
homogeneous diagonal models \cite{HomCosmo,Spin}. There are two aspects of 
this singularity-free property. The imposition of the Hamiltonian constraint
(``Wheeler--DeWitt equation") of LQC leads to a difference equation with 
eigenvalues of the densitized triad variable serving as labels. These 
eigenvalues can take negative values corresponding to reversal of orientation.
The difference equation, viewed as an evolution equation in these labels, 
allows solutions to evolve through the zero eigenvalue (zero size). Thus there
is no breakdown of evolution equation at the classically indicated singularity
at zero size. This is the first aspect of absence of singularity. The second
aspect is that matter densities and curvatures remain finite at all
sizes. The inverse scale factor operator that enters the definitions of
these quantities turns out to have bounded spectrum. For an explanation and
details see \cite{Bohr,InvScale}.
%
%is not the inverse of the scale factor but is another
%operator which has a bounded spectrum and whose eigenvalues are inverses
%of eigenvalues of the scale factor operator for large eigenvalues
%\cite{InvScale}.

While quantum theory is well specified at the kinematical level, one
still does not have a physical inner product to have a bona-fide Hilbert
space of solutions of the Hamiltonian constraint (the difference
equation). The issue of Dirac observables is also under-explored (but
see \cite{GolamHubble}). Consequently, understanding of semi-classical 
behavior in terms of expectation values of observables is not yet available.
To relate implications of LQC which is based on a discrete quantum geometry,
to observable (and more familiar) quantities described in terms of the
continuum geometrical framework of general relativity, the idea of an
{\em effective Hamiltonian} has been proposed \cite{SemiClass,Chaos}. 
This Hamiltonian contains the modifications implied by LQC to the usual 
classical Hamiltonian. This approach retains the kinematical framework of 
Robertson-Walker geometry but gives modification of the dynamics of the 
Einstein equations.

The effective Hamiltonian in LQC is generally derived in two steps 
\cite{Chaos,EffectiveHamiltonian}. In first step, one develops a 
{\em continuum approximation} to the fundamental difference equation to 
obtain a differential equation \cite{Fundamental}. For large volumes 
where one expects the manifestation of discrete geometry to be negligible, 
the differential equation matches with the usual Wheeler-DeWitt equation 
(with certain factor ordering). In the second step one looks for a WKB form 
for solution of the differential equation to derive the corresponding 
Hamilton-Jacobi equation and read-off the Hamiltonian. This is the effective
Hamiltonian. The effective Hamiltonian differs from the classical
Hamiltonian due to the modifications in the differential equation derived
from the difference equation. There are two sources of modifications. In
the matter sector, the modifications come from using the modified
inverse triad operator which incorporate the small volume deviations.
These involve inverse powers of the Planck length and thus are
non-perturbative. One can also get modifications in the gravity sector for
small volumes. These have been obtained recently \cite{EffectiveHamiltonian},
exploiting non-separable structure of the kinematical Hilbert space of LQC 
\cite{Bohr}. 

It turns out that the dynamics (evolution with respect to the synchronous
time) implied by the effective Hamiltonian captures essential features of 
the difference equation, in particular the dynamics is {\em non-singular}. 
A universe beginning at some large volume will never reach zero volume when 
evolved backwards. Since the framework for effective dynamics is that of 
the usual pseudo-Riemannian geometry, the arguments leading to the singularity
theorem are applicable and therefore non-singular evolution must imply 
violation of the strong energy condition on effective matter density and 
pressure. While the effective density and pressure \cite{EffectiveHamiltonian}
includes contributions from gravity sector alone, in this letter we concentrate
on the matter sector modifications only.

The question we address is whether the modifications in the matter sector 
imply violation of strong energy condition. In general the strong energy
condition requires $ R_{\alpha\beta}\xi^{\alpha}\xi^{\beta}=
8 \pi G (T_{\alpha\beta} - \frac{1}{2}g_{\alpha\beta} T)
\xi^{\alpha}\xi^{\beta} \geq 0$, for all time-like vectors $\xi^{\alpha}$.
Within the context of homogeneous and isotropic geometries, the strong 
energy condition applied to the congruence of isotropic observers (or
four velocity of the matter perfect fluid), becomes $R_{00} = 4 \pi G (\rho
+ 3 P) \ge 0$ where $\rho$ is the total energy density and $P$ is the total 
pressure of the matter fluid. Defining $\omega := P/\rho$ (with $\rho $ 
assumed to be positive definite) as the equation of state variable, the 
violation of strong energy condition is conveniently stated as $\omega < 
- \Case{1}{3}$. Note that since $R_{00} = - \Case{\ddot{a}}{a}$, violation 
of the strong energy condition in this context also implies an 
{\em accelerated} evolution of the scale factor or in other words an 
{\em inflationary phase}.

For simplicity, let the matter sector consists of a single scalar field with
a standard kinetic term and self interaction potential. For spatially
homogeneous and isotropic fields, the density and pressure are read-off 
from the perfect fluid form as $\rho = \Case{1}{2} \dot{\phi}^2 + V(\phi)
~,~P = \Case{1}{2} \dot{\phi}^2 - V(\phi)$ while the classical Hamiltonian
is $H_{cl} =  \frac{1}{2} a^{-3} {p_{\phi}}^2 + a^{3} V(\phi)$, where 
$p_{\phi}$ is the conjugate field momentum. Note that in LQC, the scale 
factor has dimensions of length.
%(In LQC, densitized triad is redefined to absorb the coordinate
%volume integral, making the scale factor $a^2 =: |p|$ to have dimensions
%of length. 
%This is to be kept in mind while making comparison with the
%standard scenarios which use a dimensionless scale factor.)
The LQC modifications are incorporated by replacing the $a^{-3}$ by a 
function coming from the definition of the inverse triad operator. The so
modified effective matter Hamiltonian is then given by $ H^{\text{eff}} =
\frac{1}{2} {|\tilde{F}_{j,l}(a)|}^{\frac{3}{2}} {p_{\phi}}^2	+
a^{3} V(\phi)$, where $\tilde{F}_{j,l}(a) = (\frac{1}{3}\gamma\mu_0 j
l_p^2)^{-1} F_l( {(\frac{1}{3}\gamma\mu_0 j l_p^2)}^{-1} a^2 )$ \cite{Refs}. 
The $j$ and $l$ are two quantization ambiguity parameters
\cite{Ambig,ICGCAmbig}. The half integer  $j$ corresponds to the dimension 
of representation while writing holonomy as multiplicative operators while
the real valued $l$ ($0<l<1$) labels different, classically equivalent
ways of writing the inverse power of the scale factor in terms of Poisson
bracket of the basic variables. A smooth approximation (except at one point) 
to the function $F_l(q)$ is given by \cite{Chaos}
\begin{eqnarray}
F_l(q)&:=& \left[ \frac{3}{2(l+2)(l+1)l} \left( ~
(l+1) \left\{ (q + 1)^{l+2} - \right. \right. \right. \nonumber \\
& & ~~ \left. |q - 1|^{l+2} \right\} ~-~ 
(l+2) q \left\{ (q + 1)^{l+1} - \right. \nonumber \\
& & ~~ \left. \left. \left. \sgn(q - 1) |q - 1|^{l+1} \right\}
~ \right) ~\right]^{\frac{1}{1-l}} \nonumber \\
& \rightarrow &  q^{-1}  ~~~~~~~~~~~~~~~~~(q \gg 1) \nonumber \\
& \rightarrow &  \left[ \frac{3 q}{l+1}\right]^{\frac{1}{1-l}}
~~~~~~~(0 < q \ll 1) ~. 
\label{invfun}
\end{eqnarray}
Thus, for the large values of the scale factor one has the
expected classical behavior for the inverse scale factor and the
quantum behavior is manifested for small values of the scale
factor. 

The density and pressure are usually defined from the perfect fluid form 
of the usual stress tensor which in turn is derived from an action principle
for the matter field. Since the LQC modifications are incorporated at
the level of the Hamiltonian, this route is not available. It is possible
to define the density and pressure directly in terms of the matter Hamiltonian. 
%This can be done by writing the Hamilton's equations (for matter and gravity)
%and recasting them in the form of the usual Raychoudhuri and Friedmann 
%equations to read off the density and pressure. 
This has been done generally in \cite{EffectiveHamiltonian} (See also
\cite{OsciUniv}). The relevant definitions, in terms of the notation
in \cite{EffectiveHamiltonian}, are ($'$ denotes $\Case{d}{d a}$)
\begin{equation}
\rho \ = \ \frac{32}{3} \frac{\alpha}{a^4} H ~,~ 
P \ = \ \frac{32}{9} \frac{\alpha}{a^4} \left\{ \left( 1 - \frac{a
\alpha'}{\alpha}\right)H - a  H'\right\} 
\label{DensityPressure}
\end{equation}
In the above $\alpha$ is a specific function of $a$. For large $a^2$, $\alpha$ 
goes as $\Case{3}{4}a$ giving the familiar form of density as $8 Ha^{-3}$
\cite{EffectiveHamiltonian}. For the LQC modified scalar matter Hamiltonian, 
and for large $a$, these definitions of density and pressure match with
those of \cite{OsciUniv}. The conservation equation $a \rho' = -3 (\rho +
P)$ is of course, automatically satisfied.  It is straight forward to
verify that for the classical Hamiltonian for the scalar field, these
reduce to the usual definitions for large $a$. 

Consider the two equation of state variables, $P/\rho$, defined by the 
effective Hamiltonian and by the classical Hamiltonian. 
\begin{eqnarray} 
\omega^{\text{eff}} & = & 
- \frac{ \Case{1}{4} p^2_{\phi} a^{-3} [\tilde{F}_{j,l}(a)]^{\Case{1}{2}}
(a [\tilde{F}_{j,l}(a)]' ) + V(\phi)}
{\Case{1}{2} p^2_{\phi} a^{-3} [\tilde{F}_{j,l}(a)]^{\Case{3}{2}} + V(\phi)}
\nonumber \\
& & + \frac{1}{3}\left(1 - \frac{a \alpha'}{\alpha}\right) 
\label{OmegaEffDef}\\
\omega & = &  
\frac{ \Case{1}{2} p^2_{\phi} a^{-6} - V(\phi)}
{ \Case{1}{2} p^2_{\phi} a^{-6} + V(\phi)} \label{OmegaDef}
\end{eqnarray}
The second term in eq.(\ref{OmegaEffDef}) vanishes for large volumes and
goes to $- 1/3$ for small volumes \cite{EffectiveHamiltonian}. It is
independent of the matter variables and will be suppressed below (see
remarks on the scales at the end).  The dynamical evolutions of these
equations of state is of course governed by the corresponding Hamiltonians.
It is however possible to derive qualitative behavior of
$\omega^{\text{eff}}$ for small scale factors {\em without having to know
explicit time evolution}, as follows. 

The equations (\ref{OmegaEffDef}, \ref{OmegaDef}) can be thought of as
two homogeneous algebraic equations for $p^2_{\phi}, V(\phi)$. For
non-trivial values of these, the determinant must vanish which gives a
relation between the two $\omega$'s as,
\begin{equation}
\omega^{\text{eff}} =  -1 + \frac{(1+\omega)a^3
[\tilde{F}_{j,l}(a)]^{\Case{3}{2}} \left(1 - \Case{a [\tilde{F}_{j,l}(a)]'}
{2 \tilde{F}_{j,l}(a)}\right)}
{(1+\omega)a^3 [\tilde{F}_{j,l}(a)]^{\Case{3}{2}}  + (1-\omega)} ~.	
\label{EffectiveEOS}
\end{equation}
Using the expression (\ref{invfun}) it is easy to see that for
the large values of the scale factor $a$, where one expects the
quantum effects to be small, $\omega^{\text{eff}} = \omega$ and the
dynamical evolution is controlled by the classical Hamiltonian. However
for small values of $a$ the $\omega^{\text{eff}}$ differs from the
classical $\omega$ dramatically. 

The numerator in the second terms of (\ref{EffectiveEOS}) vanishes as
$a^{3 + \Case{3}{1 - \ell}}$. If $ 1 - \omega$ in the denominator
dominates, then clearly $\omega^{\text{eff}} \to -1$. This would happen
either because $1 - \omega \nrightarrow 0$ or it vanishes slower than 
$a^{3 + \Case{3}{1 - \ell}}$. In the former case we already have
violation of strong energy condition. It is possible to get constraints on 
the behavior of $\omega$ as the scale factor vanishes. For instance,
the conservation equation expressed in terms of the scale factor implies
that if $\omega \to 1$ then $\rho \sim a^{-6}$. This equation is
independent of the LQC modification and applies also to effective
density. Furthermore, from the definition it follows that $ 1 - \omega = 
\Case{2 V(\phi)}{\rho} $. Thus, the $1 - \omega$ term in the denominator
will dominate if $V(\phi(a))\ a^{- \Case{3 \ell}{1 - \ell}}$ diverges as
$a \to 0$. This dominance is ensured if either (i) the potential never vanishes 
during the evolution or (ii) $V(\phi(a))$ vanishes at the most as a power law,
$a^{\xi}$. In the former case, $\omega^{\text{eff}} \to -1$ will hold
independent of the ambiguity parameter $\ell$ while in the latter case,
for any given $\alpha$ we can always {\em choose} $\ell >
\Case{\xi}{\xi + 3}$ so that $\omega^{\text{eff}} \to -1$ is achieved.
Note that this is {\em not} a fine tuning.

\begin{figure}[htb]
\begin{center}
\includegraphics[width=8cm,height=6cm]{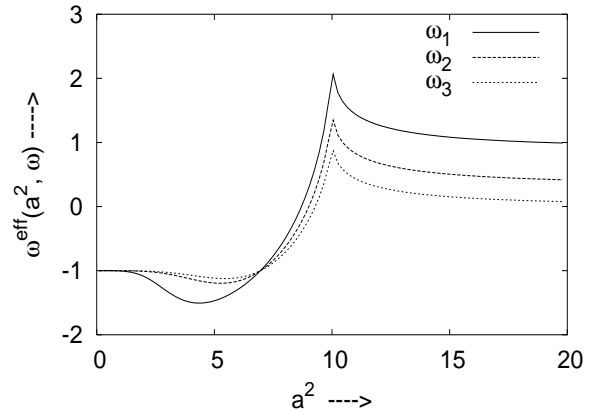}
\end{center}
\caption{Plot of $\omega^{\text{eff}}$ as a function of $a^2$ and $\omega$ for
different constant values $\omega = 0.9, 0.33, 0.0001$. The ambiguity
parameters are $j = 5, l = 0.5$ and $a^2$ is in units of
$\Case{1}{6}\gamma\mu_0\lP^2$. For small scale factor, $\omega^{\text{eff}}$ 
always approaches $ -1$ from below while for larger values it approaches
$\omega$.}
\label{sizefig}
\end{figure}

For the special case of identically vanishing potential, we get $\omega = 1$
and the expression for $\omega^{\text{eff}}$ simplifies to 
$- \Case{a [\tilde{F}_{j,l}(a)]'} {2 \tilde{F}_{j,l}(a)}$. For small
scale factor $\omega^{\text{eff}} \to - \Case{1}{1 - \ell} \ < -1$ and violation
of strong energy condition follows. Indeed, since $\omega^{\text{eff}}<-1$ 
holds,
one has a phase of {\em super-inflation}. In fact this feature corresponds 
to situation considered in \cite{Inflation,ParamShinMart}. However this feature
is rather special because even a tiny but non-negative potential will force 
$\omega^{\text{eff}}$ to take the form (\ref{EffectiveEOS}) (see
figure[\ref{sizefig}]).

The spikes in the figure correspond to the non-differentiability of the
$F_l(q)$ at $q = 1$. These can be removed by a local smoothing of the function
around $q = 1$ and thus have no physical significance.'

In summary, we find that if the scalar field potential satisfies $V(\phi)
> 0$ then irrespective of what values we choose for the ambiguity parameters
and irrespective of `initial conditions' for the scalar field, there is
always a violation of strong energy condition in the small volume regime
and of course a corresponding inflationary phase. Furthermore since the
effective equation of state variable approaches $- 1$, we get to a phase
of exponential inflation. If the potential has zeros which are
approached as a power law for small scale factor, one can always choose
a value of $\ell$ to get the same result. We emphasize that unlike the 
usual inflationary scenarios we do {\em not} need to invoke `slow roll 
conditions' which constrain the potential as well as initial conditions 
for the scalar and effectively posit the equation of state variable to be 
$- 1$. It is enough to have the evolution get to small volume regime to 
generate (exponential) inflation. 

%The predictions such as the {\em scale-invariant} primordial power spectrum
%of standard of inflationary scenario may then be expected to be true in this
%LQC generated inflation as well. The details of the primordial density 
%fluctuation, including the effect of {\em zero-point} proper length
%\cite{GolamPrep} are under investigation and will be published else where.

A couple of remarks are in order. Firstly, if LQC modifications from
gravity sector (quantum geometry potential) are also included 
\cite{EffectiveHamiltonian}, then the results regarding behavior of the
effective equation of state as a function of the scale factor, are unchanged. 
These gravitational contributions to density and pressure violate the strong
energy condition by themselves. Their effective equation of state parameter 
is $+ 1$ but both the density and pressure are negative. 

%{\bf Where is bounce scale?}

A second remark concerns the scales. There are two basic scales available: (i)
the `quantum geometry scale', $L_{\text{qg}}^2 := \Case{1}{6}\gamma \mu_0 \lP^2
= p_0$ \cite{Bohr} and (ii) the `inverse scale factor scale', $L_{\text{j}}^2
:= \Case{1}{6} \gamma \mu_0 \lP^2 (2j) = 2j p_0$. The former sets the scale for
non-perturbative modifications in the gravitational sector while the latter
does the same for the matter sector.  Clearly, $L_{\text{qg}} \le L_{\text{j}}$.
It is easy to see \cite{EffectiveHamiltonian} that the WKB approximation gets
poorer close to $L_{\text{qg}}$. This is consistent with the physical
expectation that below this scale one is in the deep quantum regime.
Furthermore, the effective model, including the modifications in both gravity
and matter sector, always shows a bounce i.e. a non-zero {\em minimum} scale factor
at which $\dot{a}$ vanishes \cite{GenericBounce}.  This introduces a {\em
third} scale, $L^2_{\text{bounce}}$ which is smaller than $L^2_j$. Clearly,
$L^2_{\text{bounce}} > L_{\text{qg}}^2$ must hold to remain within the domain
of validity of WKB approximation. In summary, $a \gg L_j$ is the classical
regime, while for $a < L_{\text{bounce}}$ one is strictly in the quantum domain
in the WKB sense and the effective Hamiltonian is not valid. The semi-classical
regime for the purposes of this paper has the scale factor between
$L_{\text{bounce}}$ and $L_{\text{j}}$. This implies that the suppressed term
in the eq. (\ref{OmegaEffDef}), is vanishingly small in the semiclassical
regime thus $\omega^{\text{eff}} \to - 1$.

The issue of whether the effective dynamics admits particle horizon or
not, is a separate issue. In view of the generic bounce in the effective
model \cite{GenericBounce}, the universe would have existed for infinite 
time in the past. The evolution could have been oscillatory or there could 
have been just one bounce in the past. In the large volume regime, we have 
the usual decelerating evolution (modulo $\Lambda$-term) implying that the
scale factor will diverge at the most as linear power of the synchronous time.
For both possibilities, the conformal time integral would be infinite implying
absence of particle horizon. However, {\em independent of the non-existence 
of particle horizon}, inflation comes built-in with the LQC modifications. 

\begin{acknowledgments}
We would like to thank Martin Bojowald for critical comments.
\end{acknowledgments}

%%%%%%%%%%%%%%%%%%%%%%%%%%%%%%%%%%%%%%%%%%%%%%%%%%%%%%%%%%%%%%%%%%%%%%%

\end{document}